# Proving False in Object-Oriented Verification Programs by Exploiting Non-Termination

Jaymon Furniss
School of Engineering and
Computer Science
Victoria University of Wellington
Wellington, New Zealand
jaymonfurniss123@gmail.co.nz

## ABSTRACT

We looked at three different object-oriented program verifiers: Gobra, KeY, and Dafny. We show that all three can be made to prove false by using a simple trick with ghost variable declaration and non-terminating code. This shows that verifiers for these languages can produce unsound results without much difficulty and that this is possibly common throughout all OO verifiers.

## 1 Introduction

This paper aims to take a brief look into a way to show that verifiers for object-oriented programming languages are unsound. The main culprit for this unsoundness is the difficulty of verifying the termination of programs written in OO languages. Out of the three languages we looked at, we were able to prove false in all of them by using a simple technique that exploits undetected non-termination.

## 2 Breaking Gobra

The first verifier we will look at is Gobra[1] [1], which is made for verifying programs in the Go programming language. Gobra is a verifier that claims to provide partial correctness verification for code written in Go. They described partial correctness as checking for correct behaviour but not for termination [2]. This means that for the verifier to be sound it should correctly verify the behaviour of the program if it successfully terminates.

```
1    package test
2    ensures false
3    func recurse() (_ int) { return
     recurse() }
4    ensures false
5    func test() { ghost var _ = recurse() }
```
*Explanation.* [2]

In lines (2-3) we define the function that promises to return an `int` and simply calls itself. It's sufficient for us to just ensure that `false` holds inside this function since it proves automatically. This is because the only thing that happens inside `recurse()` is that it calls `recurse()` again, therefore it uses its own post-condition to prove itself. In an environment where only partial correctness is considered, this is perfectly sound since this function would never terminate and thus no contradictions would hold during runtime.

Lines (4-5) define the function `test()` that breaks this soundness by using a ghost variable assignment. Using a ghost expression here means that this line of code is only considered by the verifier and is omitted during runtime. We exploit this to create a variable that, if assigned, must prove `false`. In a runtime environment, this would loop forever and never happen, but the ghost expression allows us to remove this recursive behaviour and keep the contradiction. We are left with a program that is verified by Gobra that terminates but verifies post-conditions that should not hold. Consider the code below as an example:

```
1    ensures r == 0
2    func bad() (r int) {
3      test()
4      return 1
5    }
```

The post-condition that 0 is returned is obviously false since the function returns 1, but by using our `test()` method from above we cause Gobra to verify this clearly incorrect behaviour. It is for this reason that we believe that allowing ghost code with a verifier that does not check for termination is unsound.

---

[1] Gobra version used is from the repository https://github.com/viperproject/gobra commit 693b2f01638afdaacbd74f1250b8b5d7b74bf866

[2] This code can be run using `java -jar -Xss128m gobra.jar -i [filename]`

## 2 Breaking KeY

KeY[3] [3] is a verifier for Java programs that has measures in place to check for termination but these measures are still far from perfect.

```
1    public class Test {
2      //@ ensures false;
3      public int recurse1() {
4        return recurse2();}
5      //@ ensures false;
6      public int recurse2() {
7        return recurse1();}
8      //@ ensures false;
9      void test() {
10       //@ ghost int x = recurse1();
11       return;}
12   }
```

*Explanation.*

Lines (2-7) define two methods, `recurse1()` and `recurse2()`. Both of these methods are very similar to the `recurse()` function in the Gobra example and behave as if you duplicated the original and made each duplicate call the other instead. Both methods ensure that false holds, and both contracts for these methods verify. KeY does check if a method depends on itself to verify, meaning the approach used for Gobra doesn't work, but it fails to check if its dependencies have circular references back to itself. This is because if a method is annotated with post-conditions, it will look no further and just assume that they hold while ignoring the possibility of circular dependence. This example shows how simply just looking at pre- and post-conditions of methods for verification is unsound in this setting.

Lines (8-12) define the exact same behaviour as the previous Gobra example. The post-condition `false` is proved by creating a dangerous ghost variable with an infinitely recursive assignation. We are again left with a program that will terminate but can also assert contradictions.

## 2 Breaking Dafny

Dafny[4] [4] is a standalone language with verification that is by far the least easy to break out of the three looked at in this paper. It is still possible to fool Dafny with non-terminating code though, so the same result is still achievable.

```
1    trait Uninhabited {
2    function method get(): int ensures
     false}
3    class Omega {
4      const omega: Omega -> Uninhabited
5      constructor() {
6    omega := (o: Omega) => o.omega(o);}
```

```
7    }
8    method test() ensures false {
9      var o := new Omega();
10     ghost var _ := o.omega(o).get();
11   }
```

*Explanation*:[5]

Lines (1-2) define the trait `Uninhabited` which contains one abstract function, `get()`, which returns an int. We have given the method the post-condition false, which is what we will use to verify another false post-condition later in the code. Under normal circumstances, there should be no way to instantiate `Uninhabited` because we would need to override its method with a body that somehow proves false.

Lines (3-7) define the `Omega` class. `Omega` declares one constant which is a function from `Omega` to `Uninhabited`. The constructor of `Omega` then assigns the omega combinator as the function `omega()`. This is the simplest way to create recursion without the function directly calling itself (writing 'this.omega(o)' doesn't work, for example). Doing this is sufficient to stop the verifier from noticing that there is possible recursion. Assigning this directly, as shown below, would result in a verification error caused by recursive dependency.

```
1    class Omega {
2      const omega: Omega -> Uninhabited :=
     (o: Omega) => o.omega(o);
3    }
```

In line (9), we declare the variable `o`, an instance of `Omega`. In line (10) we use `o` to create an instance of `Uninhabited` which we use to call `get()` and receive an `int`. This, of course, should not work because the program would loop forever before instantiating an instance of `Uninhabited`, but the verifier does not know this. Dafny then assumes that a value is assigned, thus assuming that the post-condition `false` in `get()` holds, ultimately achieving the same result as the previous two examples.

## 5 Conclusion

It is clear to see from these examples that allowing ghost code allows for unsoundness when faced with non-terminating behaviour. The Dafny example in particular shows how current verifiers have difficulty detecting possible recursion. It could be argued that using pre- and post-conditions as the basis for proving correctness makes it difficult to prove termination by abstracting away behaviour. Features like inheritance and dynamic dispatch, which were barely even needed in this study, additionally make it very easy to create hidden recursion, even by accident.

---

[3] KeY version used is 2.10.0
[4] Dafny version used is 3.8.0.40823
[5] This code can be run using `dafny /compile:3 fileName.dfy`

## ACKNOWLEDGEMENTS

I would like to thank Marco Servetto and Isaac Oscar Gariano for providing the example used to break Dafny.